# ESTIMATION OF VEGETATION LOSS COEFFICIENTS AND CANOPY PENETRATION DEPTHS FROM SMAP RADIOMETER AND ICESAT LIDAR DATA

*M. Baur[1], T. Jagdhuber[2], M. Link[2,3], M. Piles[4], D. Entekhabi[5], A. Fink[6,2]*

[1]University of Bayreuth, Institute of Geoecology, Universitätsstr. 30, 95447 Bayreuth
[2]German Aerospace Center, Microwaves and Radar Institute, P.O. BOX 1116, 82234 Wessling
[3]LMU Munich, Department of Geography, Luisenstr. 37, 80333 Munich
[4]Institut de Ciencies del Mar CSIC, Pg. Marítim de la Barceloneta 37-49, 08003 Barcelona
[5]MIT, Department of Civil and Environmental Engineering, Vassar Street 15, 02139 Cambridge
[6]University of Augsburg, Institute of Geography, Alter Postweg 118, 86159 Augsburg
Email: martin.baur@uni-bayreuth.de, {thomas.jagdhuber, moritz.link}@dlr.de, mpiles@icm.csic.es, darae@mit.edu, anita.fink@student.uni-augsburg.de

## ABSTRACT

In this study the framework of the $\tau$ - $\omega$ model is used to derive vegetation loss coefficients and canopy penetration depths from SMAP multi-temporal retrievals of vegetation optical depth, single scattering albedo and ICESat lidar vegetation heights. The vegetation loss coefficients serve as a global indicator of how strong absorption and scattering processes attenuate L-band microwave radiation. By inverting the vegetation loss coefficients, penetration depths into the canopy can be obtained that is displayed for the global forest reservoirs. A simple penetration index is formed combining vegetation heights and penetration depth estimates. The distribution and level of this index reveal that for densely forested areas the soil signal is attenuated considerably, which can affect the accuracy of soil moisture retrievals.

## 1. INTRODUCTION

As a very fundamental assumption all soil moisture retrieval algorithms are dependent on a sufficient soil signal after radiating trough the vegetation canopy [1]. Within the widely used $\tau$ - $\omega$ model [2], a zeroth order solution of the radiative transfer equation, the attenuation due to the vegetation canopy is expressed using the vegetation optical depth $\tau$ and the single scattering albedo $\omega$. Besides the quantification of canopy attenuation both parameters can have biophysical importance and can be empirically linked to vegetation characteristics, like vegetation biomass [3] or vegetation water content [4]. In the following the $\tau$ - $\omega$ model will be used as a basis to derive vegetation loss coefficients and penetration depths into canopy. In a next step global L-band penetration will be calculated and analyzed.

## 2. DEFINITION OF VEGTATION LOSS COEFFICIENTS AND PENETRATION DEPTH

Generally microwave radiation within a vegetation medium is attenuated in proportion to the mediums extinction coefficient $K_e$ [$m^{-1}$], which is composed out of scattering and absorption losses, represented by their loss coefficients $K_s$ [$m^{-1}$] and $K_a$ [$m^{-1}$]. Hence the extinction coefficient is defined as [5, 6],

$$K_e = K_s + K_a. \qquad (1)$$

$\omega$ and $\tau$ can be derived using the loss coefficients. The single scattering albedo $\omega$ [-] is derived from the scattering loss relative to the total extinction [7],

$$\omega = \frac{K_s}{K_e} = \frac{K_s}{K_s + K_a}. \qquad (2)$$

The vegetation optical depth $\tau$ [-] can be linked to $K_e$ in the following manner [8],

$$\tau = K_e \cdot h, \qquad (3)$$

with *h* [m] being the layer thickness, or vegetation height. This quantity describes total attenuation for a vegetation layer, dependent on $K_e$ and *h*. Following definitions in (1) to (3) all loss coefficients can be derived as:

$$K_e = \frac{\tau}{h}, \qquad (4)$$
$$K_s = \frac{\tau}{h} \cdot \omega, \qquad (5)$$
$$K_a = K_e - K_s = \frac{\tau}{h}(1 - \omega). \qquad (6)$$

Inversion of the vegetation loss coefficients leads to vegetation/canopy penetration depth $\varrho$, a depth at which the intensity of the electromagnetic wave has fallen to *1/e* of its initial intensity due to attenuation within the vegetation layer

[9]. Canopy penetration depths $\varrho_{K_e}$ [m], $\varrho_{K_s}$ [m], $\varrho_{K_a}$ [m] can be calculated for all three loss coefficients:

$$\varrho_{K_e} = \frac{1}{K_e}, \quad (7)$$
$$\varrho_{K_s} = \frac{1}{K_s}, \quad (8)$$
$$\varrho_{K_a} = \frac{1}{K_a}. \quad (9)$$

Equation (9) and (10) indicate rather special cases of penetration depth compared to (8), as attenuation normally is caused by both, scattering and absorption, within the canopy and not only by one of the two. Dividing $\varrho_{K_e}$ by $h$ can give an estimate (index), if the microwave signal penetrates through the vegetation layer to the soil, or if the signal decays within the canopy due to attenuation.

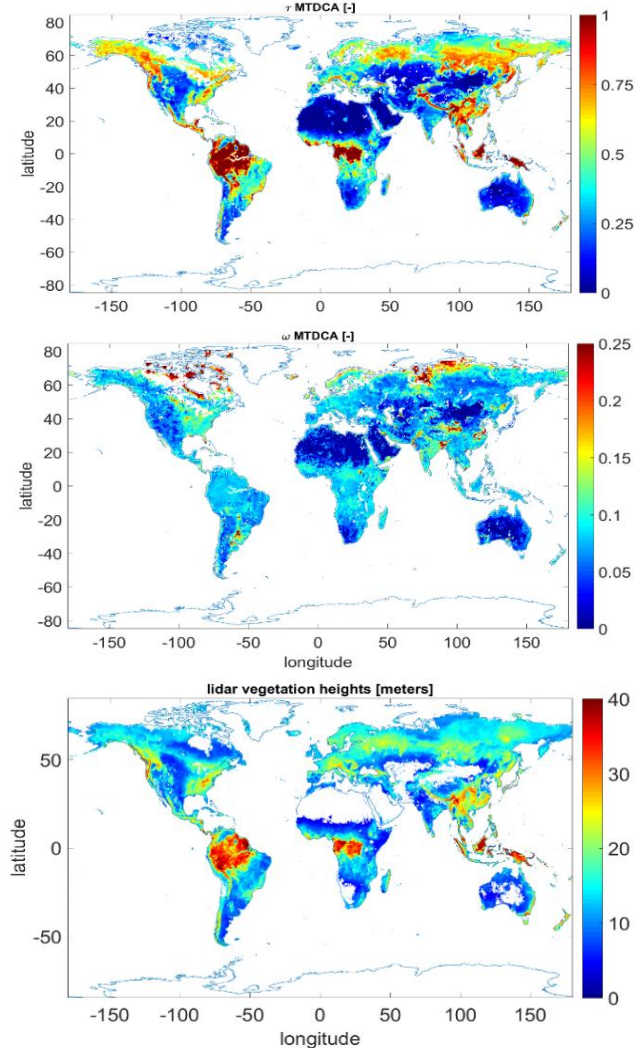

Fig. 1. Global input data. $\tau$ and $\omega$ derived from SMAP time series using an MT-DCA algorithm and lidar vegetation heights $h$ from ICESat lidar sensor.

$$Index\ of\ penetration\ [-] \to \frac{\varrho_{K_e}}{h} = \frac{1}{\tau} \quad (10)$$

Index of penetration values lower than one indicate that the signal is attenuated within the vegetation layer to an intensity below *1/e*. Values higher than one indicate penetration through the vegetation layer. This index cannot give a definite answer, if a sufficient signal for the retrieval of soil parameters is available, but indicate that values might be spurious and should be quality flagged.

### 3. ESTIMATION OF LOSS COEFFICIENTS FROM SMAP AND ICESAT DATA

Both $\tau$ and $\omega$ can be estimated from satellite data using different retrieval approaches. In this case they were retrieved from SMAP radiometer data using a multi temporal dual channel algorithm (MT-DCA) [10]. Lidar vegetation heights were accessed from ICESat GLAS sensor [11]. They were processed for the SMAP active-passive period from 13.04. to 07.07.2015 and averaged over this time period. The input parameters for loss and penetration depth estimation are shown in Fig. 1. $\tau$ has highest values for tropical forest and shows accordance with vegetation height. $\omega$ indicates high values in artic regions, what can be influenced by freeze-thaw dynamics, and in river deltas. Fig. 2 shows global plots of mean vegetation loss coefficients $K_e$, $K_s$ and $K_a$. All three parameters have different spatial patterns, where $K_a$ is mainly driven by $K_e$, as it is one order of magnitude bigger than $K_s$. The extinction coefficient $K_e$ generally shows high values for boreal forest areas. In addition to that there are several extinction hotspots e.g. the Gran Chaco, Southern Horn of Africa and Yucatán peninsula. All three parameters are dependent on seasonal dynamics and change of vegetation cover (especially crop cycle) and therefore only valid for the active-passive period of the SMAP mission. For further analysis of the presented methods, C-band data from AMSR-E could be used to compare vegetation loss/attenuation results from different frequencies. This can give further insights on the frequency dependence of the loss coefficients and eventually $\omega$ and $\tau$.

### 4. PENETRATION DEPTHS AND A PENETRATION INDEX FOR GLOBAL FOREST AREAS

Vegetation loss coefficients were inverted to estimate the penetration depth into the vegetation canopy. Penetration depths have to be used with care, as they assume the target coefficients to be fully representative and constant for the entire attenuating vegetation layer. Hence penetration depths should have a higher representativeness for vegetated areas with sufficiently closed canopy. Therefore penetration depths in Fig. 3 were only calculated for the global forest areas indicated by the IGBP landcover classes 1-5 (evergreen needleleaf – evergreen broadleaf – deciduous needleleaf – deciduous broadleaf – mixed forest). Fig. 4 shows the distributions of the penetrations depths as boxplots. Considering the IGBP classes, all show fairly similar values, with deciduous broadleaf forests having the biggest spread

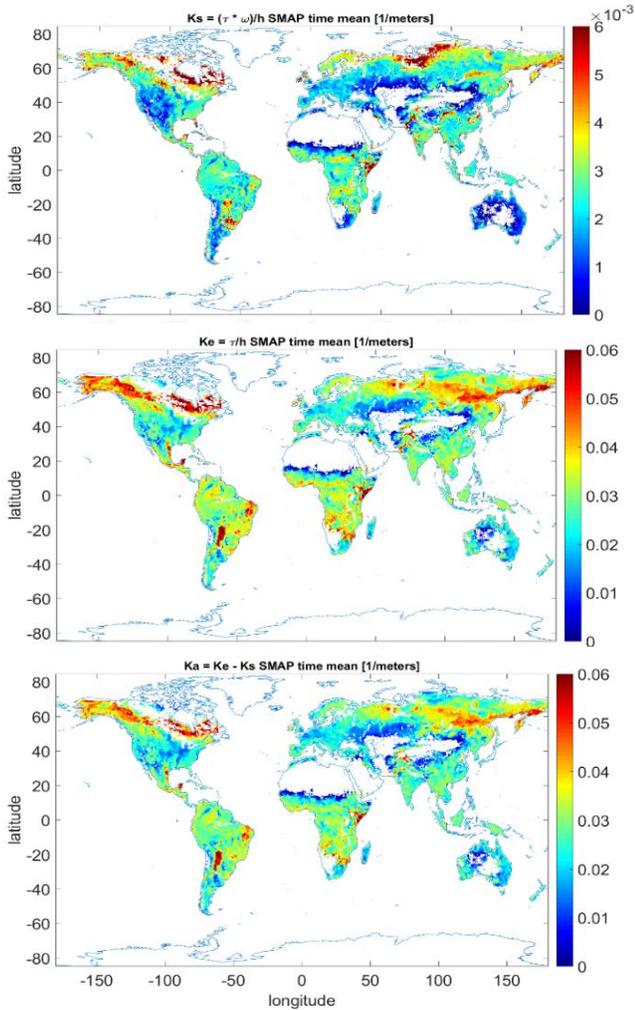

Fig. 2. Mean vegetation loss coefficients $K_s$, $K_e$ and $K_a$ calculated from SMAP vegetation optical depth ($\tau$), single scattering albedo ($\omega$) and lidar vegetation heights $h$. The mean was calculated for SMAP active-passive period.

Distinct growth stages during the acquisition time could be a possible reason. Dividing $\varrho_{K_e}$ by $h$ can give an estimate (index) if the microwave signal penetrates through the vegetation layer to the soil, or if the signal is lost in the canopy due to attenuation. Values lower than one indicate that the signal is attenuated to an intensity below $1/e$ within the vegetation layer. Values higher than one indicate significant penetration through the vegetation canopy. Fig. 5 shows fairly well penetration globally, with exception of dense tropical forest. Furthermore the boxplots provided in Fig. 6 reveal concurrent results, but visualize a considerable spread for all classes with parts of the distributions having values below one (6.25% of all valid pixels show values below 1).

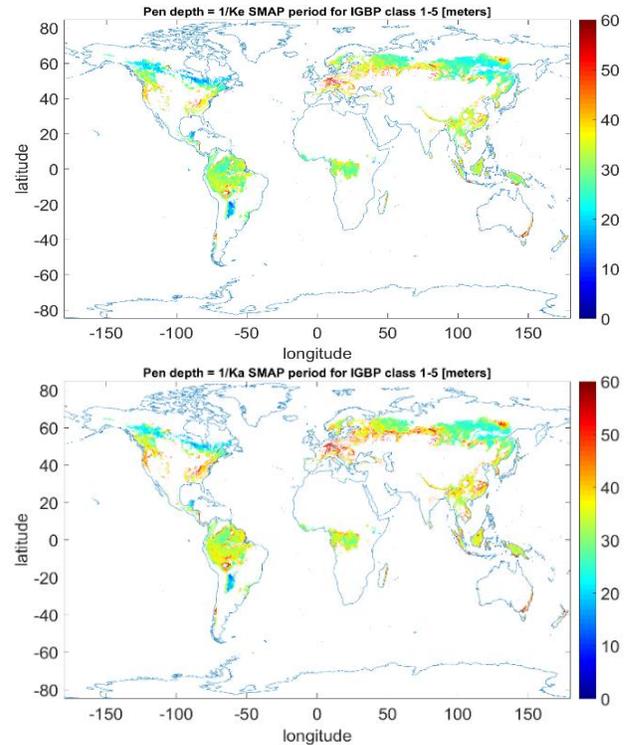

Fig. 3. Mean penetration depths $\varrho_{K_e}$ and $\varrho_{K_a}$ for IGBP classes 1-5 (forested areas), inverted from the corresponding loss coefficients. The mean was calculated for SMAP active-passive period.

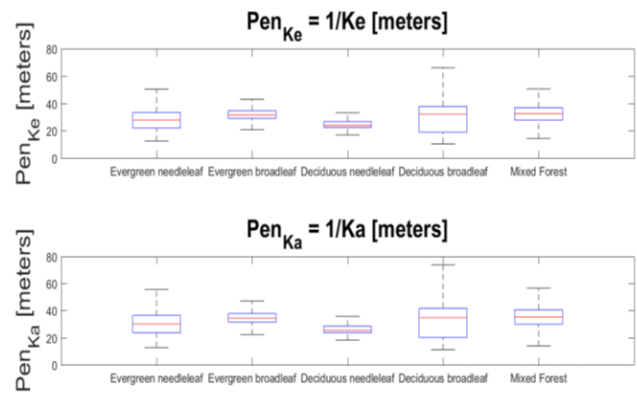

Fig.4. Box-plots showing the median, quantiles, maximum and minimum values for penetration depths $\varrho_{K_e}$ and $\varrho_{K_a}$. Within the whiskers 99.3% of the data is located, while outliers are not displayed.

Values lower than one indicate that the signal is attenuated to an intensity below $1/e$ within the vegetation layer. Values higher than one indicate significant penetration through the vegetation canopy. Fig. 5 shows fairly well penetration globally, with exception of dense tropical forest. Furthermore the boxplots provided in Fig. 6 reveal concurrent results, but visualize a considerable spread for all classes with parts of the distributions having values below one (6.25% of all valid pixels show values below 1). Nevertheless one may conclude, that at L-band even with dense vegetation cover a sufficient soil signal is obtained for most areas (93.75%)) and can be used for soil parameter retrieval purposes.

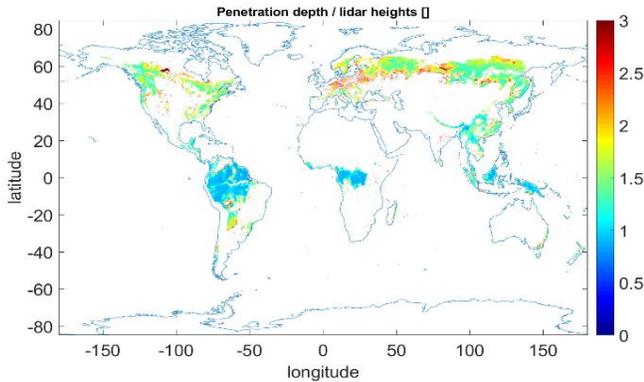

Fig. 5. Index of penetration $\frac{\varrho K_e}{h} = \frac{1}{\tau}$ for IGBP classes 1-5 (forested areas). Values above one indicate a significant penetration through the vegetation into the soil, while values below one indicate an attenuation of the signal to a value below *1/e* within the vegetation.

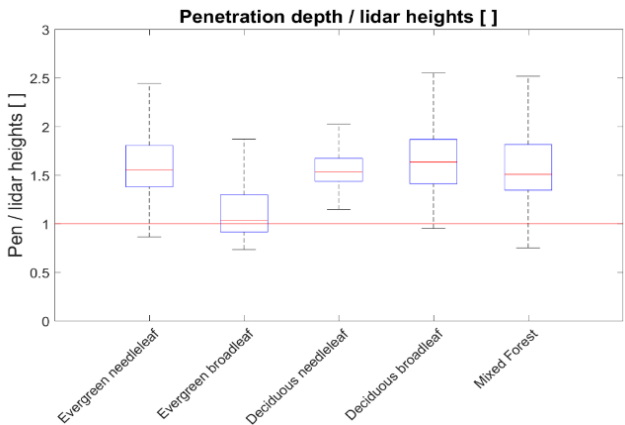

Fig. 6. Boxplots showing the median, quantiles, maximum and minimum values for the penetration index. Within the Whiskers 99.3% of the data is located, while outliers are not displayed. The red line and areas above indicate that canopy penetration to the underlying soil occurs.

## 5. REFERENCES

[1] E.G. Njoku and D. Entekhabi, "Passive microwave remote sensing of soil moisture," *Journal of Hydrology, 184*, pp. 101-129, October 1996.

[2] T.J. Jackson, T. Schmugge and J.R. Wang, "Passive microwave remote sensing of soil moisture under vegetation canopies," *Water Resources Research, 18*, pp. 1137-1142, August 1982.

[3] S. Paloscia, E. Santi, P. Pampaloni and S. Pettinato, "Multifrequency microwave emission for estimating optical depth and vegetation biomass," *2016 IEEE International Geoscience and Remote Sensing Symposium (IGARSS)*, Beijing, pp. 5296-5299, July 2016.

[4] P. O'Neill, A. Joseph, P. Srivastava, M. Cosh and R. Lang, "Seasonal parameterizations of the tau-omega model using the ComRAD ground-based SMAP simulator," *2014 IEEE Geoscience and Remote Sensing Symposium*, Quebec City, pp. 2423-2426, 2014.

[5] B.K. Hornbuckle and T.L. Rowlandson, "Evaluating the First-Order Tau-Omega Model of Terrestrial Microwave Emission," *IGARSS 2008 - 2008 IEEE International Geoscience and Remote Sensing Symposium*, Boston, pp. 193-196, 2008.

[6] J.P. Wigneron, J.C. Calvet, A. Chanzy, O. Grosjean and L. Laguerre, "A composite discrete-continuous approach to model the microwave emission of vegetation," *IEEE Transactions on Geoscience and Remote Sensing*, 33 (1), pp. 201-211, Jan 1995.

[7] P. Ferrazzoli, L. Guerriero and J. P. Wigneron, "Simulating L-band emission of forests in view of future satellite applications," *IEEE Transactions on Geoscience and Remote Sensing*, 40 (12), pp. 2700-2708, Dec 2002.

[8] T.H. Liao, S.B. Kim, S. Tan, L. Tsang, C. Su and T.J. Jackson, "Multiple Scattering Effects With Cyclical Correction in Active Remote Sensing of Vegetated Surface Using Vector Radiative Transfer Theory," *IEEE Journal of Selected Topics in Applied Earth Observations and Remote Sensing*, 9 (4), pp. 1414-1429, April 2016.

[9] M. Nolan, D.R. Fatland, "Penetration depth as a DInSAR Observable and Proxy for Soil Moisture," *IEEE Transactions on Geoscience and Remote Sensing, 41 (3), pp. 532-537, March 2003.*

[10] A. G. Konings, M. Piles, K. Rötzer, K.A. McColl, S.K. Chan, D. Entekhabi, "Vegetation optical depth and scattering albedo retrieval using time series and dual-polarized L-band radiometer observations," *Remote Sensing of the Environment,* 172, pp. 178-189, January 2016.

[11] M. Simard, N. Pinto, J.B. Fisher, A. Baccini, "Mapping forest canopy height globally with spaceborne lidar," *Journal of Geophysical Research: Biogeosciences*, 116, pp. 2156-2202, November 2011.